\documentclass[preprint,12pt]{elsarticle}

\usepackage{amssymb}

\usepackage{lineno}
\usepackage{siunitx}
\usepackage{amsmath}
\usepackage{booktabs}
\usepackage{adjustbox}
\usepackage{graphicx}

\usepackage{tikz}
\usetikzlibrary{fit,shapes.misc}

\newcommand\marktopleft[1]{%
	\tikz[overlay,remember picture] 
	\node (marker-#1-a) at (0,1.5ex) {};%
}
\newcommand\markbottomright[1]{%
	\tikz[overlay,remember picture] 
	\node (marker-#1-b) at (0,0) {};%
	\tikz[overlay,remember picture,thick,solid,inner sep=3pt]
	\node[draw,rounded rectangle,fit=(marker-#1-a.center) (marker-#1-b.center)] {};%
}

\journal{Artificial Intelligence in Medicine}

\begin{document}

\begin{frontmatter}



\title{Predicting physiological developments from human gait using smartphone sensor data}


\author{Umair Ahmed}
\author{Muhammad Faizyab Ali, Kashif Javed, Haroon Atique Babri*}

\address{University of Engineering and Technology Lahore, Pakistan}

\begin{abstract}
Coronary artery disease, heart failure, angina pectoris and diabetes are among the leading causes of morbidity and mortality over the globe. Susceptibility to such disorders is compounded by changing lifestyles, poor dietary routines, aging and obesity. Besides, conventional diagnostics are limited in their capability to detect such pathologies at an early stage. This generates demand for automatic \textit{recommender} systems that could effectively monitor and predict pathogenic behaviors in the body. To this end, we propose human gait analysis for predicting two important physiological parameters associated with different diseases, \textit{body mass index }and \textit{age}. Predicting age and body mass index by actively profiling the gait samples, could be further used for providing suitable healthcare recommendations. Existing strategies for predicting age and body mass index, however, necessitate stringent experimental settings for achieving appropriate performance. For instance, precisely recorded speech signals were recently used for predicting body mass indices of different subjects. Similarly, age groups were predicted by recording gait samples from on-body and wearable sensors. Such specialized methods limit active and convenient profiling of human age and body mass indices. We address these issues, by introducing smartphone sensors as a means for recording gait signals. Using on-board accelerometer and gyroscope helps in developing easy-to-use and accessible systems for predicting body mass index and age. To empirically show the effectiveness of our proposed methodology, we collected gait samples from \textit{sixty-three} different subjects that were classified in body mass index and age groups using \textit{six} well-known machine learning classifiers. We evaluated our system using two different windowing operations for feature extraction, namely \textit{Gaussian} and \textit{Square}. 
  
\end{abstract}


\end{frontmatter}


\section{Introduction}
\label{S:1}
Human \textit{body mass index} (BMI) is a key indicator of physiological developments within the body. Commonly known as a measure of \textit{lipid} content in an individual, it serves as a diagnostic marker for gauging tendencies towards various pathologies \cite{wolk2003body}. A seminal study, involving 310,000 participants, from 33 different cohorts, reported a compelling correlation between persistently elevated BMIs and cardiovascular diseases \cite{asia2004body}. In another investigation, Narayan \textit{et al}. \cite{narayan2007effect} suggested an elevated risk of diabetes in obese and overweight youth. More importantly, an association between BMI and cancer was also implied by an epidemiological study conducted by Renehan \textit{et al}. \cite{renehan2008body}. All such studies, signal the importance of an optimal BMI necessary for preserving physiological balance within the body. Moreover, BMI is not a static measurement of body physiology, rather, it is subject to change with \textit{aging }-- another essential determinant of human health. With aging, the obesity tendencies are known to increase, thereby altering BMIs over time \cite{wang2007forecasting,st2010body}. 
To better understand and predict the association of BMI with various diseases, it becomes necessary to profile it more frequently and automatically. This could help monitor any abormal or pathogenic variations in human BMI. Since BMI changes with progressing age, its age-sensitive nature remains relevant. Therefore, autonomous systems for predicting both age and BMI become indispensible. Estimates from such \textit{recommender} \textit{systems }could then be used for early detection and diagnosis of various pathologies including diabetes, coronary artery disease (CAD) and angina (Fig. \ref{Figure_1}). \par

Towards developing a system for predicting BMI and age groups, several methods have been developed. Traditionally, BMI is computed by its conventional definition \cite{quetelet1869physique, garrow1985quetelet} that utilizes body \textit{height} (m) and \textit{mass} (kg) of an individual:

\begin{equation}\label{BMI}
BMI = \frac{mass}{height^2}
\end{equation}

To automate BMI estimation, Lee \textit{et al}. \cite{lee2013prediction} developed a machine learning strategy wherein, they utilized carefully recorded speech signals, for predicting human BMI. These voice signals, otherwise prone to noise, were sampled under specialized experimental settings which hinder instantaneous monitoring of BMI. Moreover, common ailments like flu and fever are also known to alter human voice \cite{abaza2007effects}. Therefore, using speech signals for actively predicting BMI, appears less tractable. Similarly, age group prediction was carried out by Qaiser \textit{et al} \cite{riaz2015one}. In their approach, they utilized on-body, inertial sensors (accelerometers and gyroscopes) for evaluating age, gender and height of 26 different subjects \cite{riaz2015one}. The use of on-body sensors for recording the limb kinetics however, requires a specific setup. This again impedes automated  profiling of BMI and age of the subject. These problems with existing strategies call for a convenient, easy-to-use and accessible methodology for predicting BMI and age of different subjects. \par

In this paper, we propose human \textit{gait} as a suitable candidate for predicting both, BMI and age groups. This is because walking patterns in human beings are governed by multiple factors such as aging, body height and body mass \cite{murray1969walking,winter1991biomechanics}. These physical and physiological components contribute towards \textit{uniquely} defining the overall human \textit{gait} \cite{winter1991biomechanics,johansson1973visual,murray1964walking}. This gait pattern, inherently recurrent in nature \cite{kadaba1989repeatability}, gives rise to a \textit{personalized} walking signal which is a charactersitic manifestation of body parameters (age and BMI). Thereupon, towards eliciting tendencies towards various diseases, gait signals present a valuable resource for predicting BMI and age groups. Towards harnessing this meaningful information from gait patterns, several techniques have been developed \cite{lee2002gait,wang2003silhouette,yang2008using,kwon2014unsupervised}. Conventionally, these methods attempt to analyze gait for identifying and recognizing different activities such as walking, running and climbing. For instance, Lee \textit{et al}. processed a video sample for extracting useful gait features and devised support vector machines (SVMs) for classifying gender of various subjects \cite{lee2002gait}. Alongwith the image-based methods, sensor-based gait analysis also find its utility in recognizing ambient human activities. These methods make use of various wearable and integrated sensors (accelerometers and gyroscopes) for recording gait samples. As an example, Yang \textit{et al}. classified static and dynamic activities based on the readings from triaxial accelerometer using neural classifiers \cite{yang2008using}. Recently, on-board smartphone sensors have emerged as simpler tools for recording gait signals thereby allowing for instantaneous and accesible recording of gait patterns. Using these smartphone sensors, Kwon \textit{et al}. \cite{kwon2014unsupervised} proposed an unsupervised learning gait analysis approach for monitoring patient activities in a room. \par 

\begin{figure}
	\centering
	\includegraphics[width=0.9\linewidth]{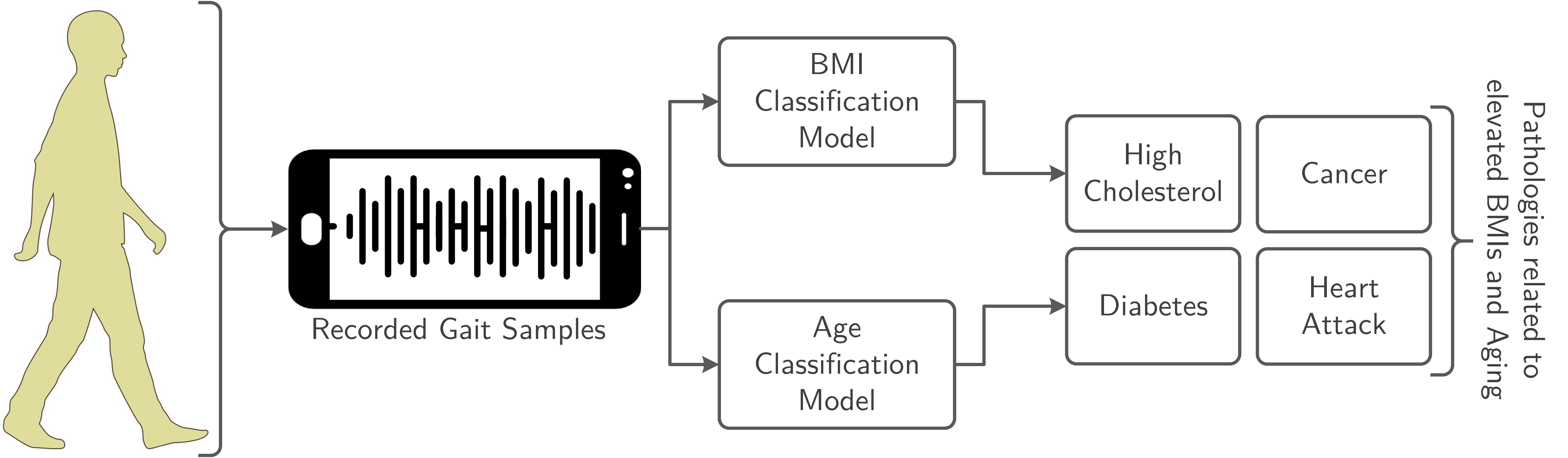}
	\caption{General theme of the proposed methodology}
	\label{Figure-1}
\end{figure}

To the best of our knowledge, gait analysis methods for predicting BMI and age groups have not been devised.
In our proposed methodology, we employ, on-board smartphone sensors for conveniently sampling the natural gait signals of various subjects (Fig. \ref{Figure_1}). We use tri-axial accelerometer to recorded linear acceleration of the body. Additionally, to capture the \textit{angular momentum }during limb activity \cite{herr2008angular}, gyroscope was also leveraged. The use of tri-axial gyroscope alongwith accelerometer, provided better coverage for the angular components of limb kinetics. Towards developing and testing the performance of our proposed technique, we recorded gait samples from $63$ different subjects. Signals from the sensors were then pre-processed and tested on \textit{six} well-known machine learning classifiers. \par
The remainder of this paper has been organized into five sections. The second section provides a review on contemporary approaches for predicting age and BMI groups. Furthermore, it highlights different methods commonly used for gait pattern analysis. In the third section, a detailed description of the proposed methodology has been described. The fourth section details experimental test-bed devised for BMI and age group classification. The fifth section presents classification results and discusses them. In the last section, we conclude the manuscript by briefly highlighting future prospects of the proposed method.

\section{Related Work}
\label{S:2}
In this section, we review various strategies commonly employed for analyzing human gait patterns. First, we present image processing based methods used for investigating different intra-limb activities. Secondly, we highlight gait analysis strategies that utilize samples  from on-body inertial sensors. Next, we discuss how smartphone sensor based methods serve as more simplistic tools for not only recording gait patterns but also for active temporal profiling of BMI and age.
\subsection{Image-based Gait Pattern Analysis Strategies}
In an earlier attempt, Shapiro \textit{et al}. \cite{shapiro1981evidence} used structural pattern recognition for studying human gait. For that, they recorded video samples from \textit{five }different runners, that were utilized for analyzing patterns of intra-limb kinetics. Similarly, Mah \textit{et al}. devised \textit{principal component }and \textit{distortion analysis }to quantitatively analyze gait under different conditions such as walking and stepping over obstacles. This helped them resolve fine differences between gait patterns in different settings. However, instead of video samples, they recorded three-dimensional joint movement data using ELITE – a digital motion analysis system \cite{mah1994quantitative,ferrigno1985elite}. Next, Lee \textit{et al}. \cite{lee2002gait} used silhouettes of gait images and trained Support Vector Machines (SVM) for classifying different persons. Noticeably, they preprocessed a video sample to extract silhouettes for further analysis. A year later, Wang \textit{et al}. \cite{wang2003silhouette} proposed a spatiotemporal gait image analysis for human identification. They employed temporally acquired silhouettes for recognition purposes. On the same lines, Xu \textit{et al}. \cite{xu2007marginal} proposed an extension to Marginal Fisher Analysis (MFA) towards processing gait images. In another important study, Yu \textit{et al}. \cite{yu2009study} employed image silhouettes for classifying gender of human subjects. \par

All aforementioned contributions indicate the utility of computer vision strategies towards recognizing human activities from patterns of their gait. Despite presence of literature highlighting image-guided gait analysis, to the best of our knowledge age and BMI prediction has not been previously undertaken.

\subsection{Wearable Inertial Sensors and Gait Pattern Analysis}
Image-based gait analyses are limited, when it comes to revealing subtle physiological states of the body. This is because images present an outward representation of the gait, hence, being unable to capture fine details underpinning inherent gait conformation. Recognizing that, wearable inertial sensors have been employed for recognizing and predicting human activity. For instance, Sabatini \textit{et al}. \cite{sabatini2005assessment} developed a gait monitoring system by recording gait samples using an integrated inertial measurement unit (IMU) including a bi-axial accelerometer and a gyroscope. They evaluated different activities such as walking, inclination and velocity from foot inertial signals \cite{sabatini2005assessment}. Later, Mannini \textit{et al}. \cite{mannini2011accelerometry} performed a comparative analysis between Hidden Markov Models (HMMs) and Gaussian Mixture Models (GMMs) for classifying human activities. Importantly, they made use of on-body accelerometers for recording gait signals \cite{mannini2011accelerometry}.  Recently, Ellis \textit{et al}. \cite{ellis2016hip} made use of accelerometers for wrists and hips to measure physical activities in humans. In an important study, Qaiser \textit{et al}. \cite{riaz2015one} used data from wearable, tri-axial accelerometer and gyroscope for predicting age, gender and height of $26$ different subjects. An important setting in their study was that IMU units were attached at four different positions on the body including chest, lower back, right wrist and left ankle \cite{riaz2015one}. \par
In such strategies, the use of wearable sensors however, makes the experiment less conducive to active monitoring of BMI and age. To this end, integrated smartphone sensors have been utilized for convenient recording gait samples. Next, we highlight these simplistic strategies that make use of smartphone sensors.
  
\subsection{Gait Pattern Analysis using Smartphone Sensor Data}
Smartphones, growing ubiquitous, have emerged as convenient tools for sampling ambulatory signals. As an example, Kwapisz \textit{et al}. \cite{kwapisz2011activity} used cell phone accelerometers for recognizing different human activities such as walking, jogging, ascending stairs, standing and sitting. In this context, Hynes \textit{et al}. \cite{hynes2009off} surveyed various mobile phone environments for recording gait signals using accelerometers. In another example, Kwon \textit{et al}. \cite{kwon2014unsupervised} proposed a simplified unsupervised learning approach for monitoring patient motion using smartphones. \par
In all such studies, smartphone based methods have mostly been used for detecting ambient human activity. We propose use of on-board accelerometers and gyroscopes from smartphones to record gait samples for predicting BMI and age groups.

\section{Methodology for Predicting Physiological Developments}
\label{S:3}
Towards developing a \textit{personalized}, easy-to-use BMI monitoring system, we devised a gait pattern analysis scheme using smartphone sensor data. For that, we leveraged on-board gyroscopes and accelerometers to record gait samples of \textit{sixty-three} different subjects. The first step involved in developing methods for predicting BMI and age group was dataset development. Gait samples from $63$ different subjects were recorded using smartphone sensors (Gyroscope and Accelerometer). We used an android application (AndroSensor) for sampling the data at 180 Hz. The raw time series data thus obtained was then pre-processed before utilizing it towards classification purposes. Data from each subject was processed to eliminate any unwanted signals recorded during the ambulatory motion. \par 
After requisite pre-processing, we extracted statistical features using two different time-series segmentation strategies. The first is Gaussian windowing operation and the second is Square or Box filter. Both of these strategies are commonly employed in literature. In our work, we utilize both of the windowing operations for comparing the performance of the these methods. Therefore, $14$ statistical features were extracted from each time series segment obtained after both Gaussian and square windowing. These $14$ features were then used for training six different classifiers. To maximize the performance of each classifier, we employed different window sizes for each classification algorithm and computed the true positive rate (TPR) against each window size. Accuracies for both Gaussian and square windows were then observed for different window sizes. This helped optimize the window parameters for different classifiers. \par 
Next, to highlight the performance of two windowing strategies, we performed principal component analysis (PCA) and observed the change in true positive rates as a result of change in number of principal components. Doing so, also helped us understand the differences in the performance of two windowing operations. Lastly, we develop a rationale for the differences in performance of the two aformentioned strategies. Figure 2 illustrates the general methodology employed towards developing classification models for predicting BMI and human age groups.

\begin{figure}
	\centering
	\includegraphics[width=0.9\linewidth]{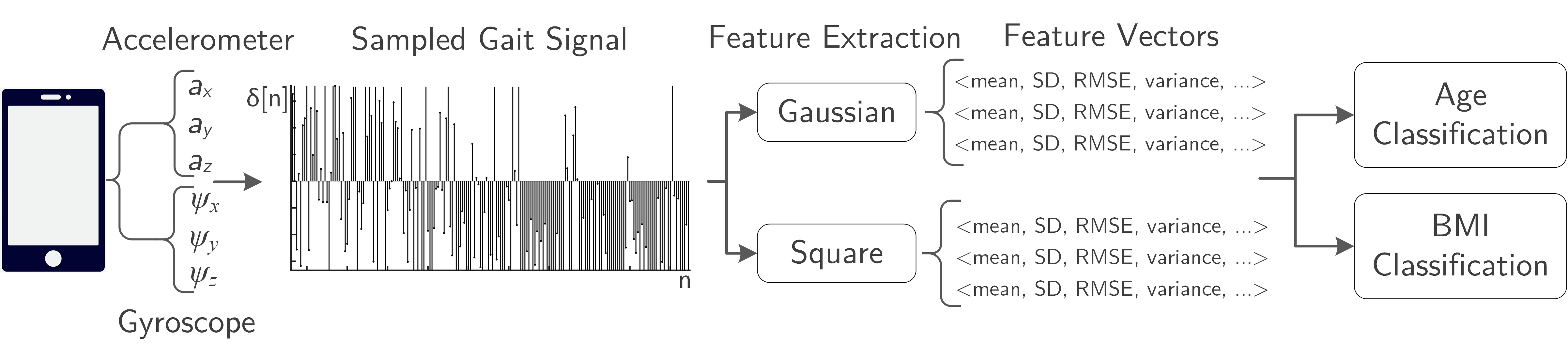}
	\caption{General workflow for predicting human BMI and age groups using smartphone sensor data}
	\label{Figure-2}
\end{figure}

\section{Experimental Testbed}
\label{S:4}
In this section, we describe the experimental settings that have been designed for performing the experiment.

\subsection{Dataset Development}
As a first step towards constructing the dataset, we recorded gait samples from \textit{eighty }different subjects. For that, we devised a simplistic data logging strategy. Instead of using wearable instruments, we leveraged two different smartphone sensors (accelerometer and gyroscope) for recording gait signals. These signals were then sampled at \SI{180}{\hertz} using an off-the-shelf android application, namely AndroSensor. Traditionally, accelerometers have been employed for recording human physical activity \cite{mannini2011accelerometry,kwapisz2011activity}. However, we make use of gyroscope as well. This is because, it provides coverage for rotational component of limb activity, thereby broadening the set of recorded signals. For consistency, gait samples were recorded by placing the \textit{same }smartphone in right pocket of each subject, who was instructed to walk over $50 ft$. Readings from tri-axial accelerometer and gyroscope  were then logged to make up \textit{raw} time-series data (Supplementary Data). Body mass, height and age for every subject was recorded in addition (Supplementary Data).\par 
The next important step in developing the dataset involved assignment of class variables to each data instance. For calculating this class information, we used the recorded height and mass of each subject. This was done by utilizing the standard definition of BMI or \textit{Quetelet Index } \cite{quetelet1869physique,garrow1985quetelet} which states:
\begin{equation}\label{BMI}
BMI = \frac{Mass}{Height^2}
\end{equation}
The calculated BMIs were then leveraged to group data subjects into various classes, as shown in Table 1. We utilized  obesity classification system devised by the World Health Organization \cite{world2000obesity} that yielded five different categories for BMI classification in Table 1. These categorical descriptions formed the basis for labeling each data record.

\begin{table}[h]
	\renewcommand{\arraystretch}{1.3}
	\caption{Categorical Description of BMI Groups}
	\vspace{0.1in}
	\label{table_bmi_groups}
	\centering
	\begin{tabular}{cc}
		\toprule[1.3pt]
		\textbf{Category} & \textbf{Class Interval (BMI)} \\ 
		\midrule[1.3pt]
		Severely Underweight &  \textless15 \\ 
		 
		Underweight & 15 - 18.5 \\ 
		
		Normal & 18.5 - 25 \\ 
		
		Overweight & 25 - 30 \\ 
		
		Severely Overweight & \textgreater30 \\ 
		\bottomrule[1.3pt] 
	\end{tabular}
\end{table}

On the same lines, we categorized \textit{sixty-three} subjects into four different age groups (Table 2). Twenty participants with ages from $10$ to $20$ were considered \textit{young}. Similarly, $20$\textit{ young adults} had ages from $21$ to $30$ while participants with ages from $31$ to $40$ were categorized as \textit{adults} ($20$ in number). Above that, $20$ \textit{aged }participants with ages between $41$ to $60$ years, Table 2.

\begin{table}[h]
	\renewcommand{\arraystretch}{1.3}
	\caption{Categorical Description of Age Groups}
	\vspace{0.1in}
	\label{table_age_groups}
	\centering
	\begin{tabular}{cc}
		\toprule[1.3pt]
		\textbf{Category} & \textbf{Class Interval (Age)} \\ 
		\midrule[1.3pt]
		Young & 10-20 \\ 
		Young Adult & 21-30 \\ 
		Adult & 31-40 \\ 
		Aged & 41-60 \\ 
		\bottomrule[1.3pt]
	\end{tabular}
\end{table}
As a whole, dataset development generated labeled, raw, time-series data from tri-axial accelerometer and gyroscopes, which was further treated before utilizing it towards developing predictive BMI and age monitoring models.

\subsection{Preprocessing}
Gait signals from tri-axial sensors can't be utilized directly for predicting BMI and age groups, since they contain extraneous values, irrelevant for analysis. To minmize the impact of such quantities, dataset was preprocessed to eliminate these superfluous factors. For that, signals were uniformly truncated to eliminate any irrelevant signal generated while placing or removing smartphone from the pocket. By doing so, we retained signal only from the gait, making it a better representative of a subject's walking pattern. The raw time-series measurements from tri-axial gyroscope  $\vec{\psi_t}$, at any time instant $t$ are generally modeled as follows:

\begin{equation}\label{GyroscopeModel}
\vec{\Psi}_t = \vec{\omega}_t + \vec{\varepsilon}_{G}
\end{equation}

where, $\vec{\Psi}_t$ is three-dimensional measurement from the gyroscopes, $\vec{\omega}_t$ is \textit{measured} angular velocity and $\vec{\varepsilon}_{G}$ is the additive Gaussian white noise, introduced as a result of inherent sensor fluctuations and electronic interferences. Similarly, three-dimensional acceleration readings from the tri-axial accelerometer $\vec{a}_t$ at any time $t$ is expressed as follows.

\begin{equation}\label{AccelerometerModel}
\vec{a}_t = \vec{g}_t + \vec{a}_{l}+ \vec{\varepsilon}_{A}
\end{equation}

where, $\vec{a}_t$ is the accelerometer reading constituted by gravitational acceleration $\vec{g}_t$, linear acceleration of the limb $\vec{a}_{l}$ and additive Gaussian white noise $\vec{\varepsilon}_{A}$. \par 
The aforementioned phase of data processing, thus yielded six raw time-series signals including, $\vec{a}_{t_x}$, $\vec{a}_{t_y}$, $\vec{a}_{t_z}$, from accelerometer and $\vec{\psi_{t_x}}$, $\vec{\psi_{t_y}}$, $\vec{\psi_{t_z}}$ from the gyroscope. Before utilizing this dataset for gait pattern analysis, we eliminated noise by extracting several useful features from the data. The next section, highlights utility of feature extraction towards developing classifcation schemes using gait samples.

\subsection{Feature Extraction}
The raw dataset, owing to its inherent stochasticity and additive white noise, is non-compliant with different classification schemes. For this reason, it becomes imperative to extract suitable features, representative of the gait signal. Statistical time-series analysis of these signals, was therefore conducted for feature extraction. We employed \textit{sliding window operation} \cite{antoniou2016digital} to divide the time-series  signal into a sequence of segments. Each segment was then utilized towards extracting several statistical features. The stochastic, non-stationary nature of gait signal makes it well-suited towards windowing operation. To avoid discontinuity in sample information, we used overlapping windows for our analysis. A suitable overlap of $50\%$ was utilized, as suggested in various literature \cite{bao2004activity,hemminki2013accelerometer,van2000shall}. We use two commonly used windowing operations, namely, \textit{Gaussian }and \textit{Square }windows. We extract features using these two different strategies to determine the most suitable for classifying BMI and age groups.  \par  
Using two different sliding windows, we extracted $14$ statistical features from each of the six time-series signals. This yielded a total of eighty-four ($84 = 6 \times 14$) features for each type of filter employed. Different window sizes for both filters were tested to identify optimal width and \textit{standard deviation} (SD) for extracting the features. This was done by varying the width (in case of Box filter) and SD (in case of Gaussian Filter). Next, we discuss Box and Gaussian filter, in relation to the sampled signals.

\subsubsection{Box (Rectangular) Windowing Operation}
Accelerometer and gyroscope generated six ($6$) different raw time-series signals, that were sampled at $180 \si{\hertz}$. Each of these signals was individually treated for extracting $14$ statistical features. Below, we discuss rectangular windowing operation leveraged to extract these features. For simplicity, let $\vec{\delta}[n]$ represent raw signal obtained from the sensor. If $\vec{\Pi}[n]$ denotes the rectangular window function, then each time-series segment $\vec{x}[n]$ obtained after windowing can be expressed mathematically as:

\begin{equation}\label{RectangularWindow}
\vec{x}[n] = \vec{\delta}(n)\vec{\Pi}(n-k\frac{W}{2}) 
\end{equation}
where, the rectangular window function, $\vec{\Pi}[n]$ with maximum window size of $W$ is defined as:
\begin{equation}
\vec{\Pi}(n)=
\begin{cases}
1 & 1\leq n \leq W \\
0 & $otherwise$
\end{cases}
\end{equation}
In Equation \ref{RectangularWindow}, $W$ is the size of square window, measured in seconds and $k=0,1,2,...$.

\subsubsection{Gaussian Windowing Operation}	
For Gaussian filtering of the dataset, a Gaussian function, $\vec{N}[n]$ is defined as:

\begin{equation}
\vec{N}[n] = \frac{1}{\sigma \sqrt{2\pi}} e^\frac{-n}{2\sigma^2}
\end{equation}

This Gaussian function was employed to segment the time-series. Therefore, if $\vec{\delta}[n]$ is raw time-series signal, the the segmented signal $\vec{x}[n]$ is obtained using Gaussian function, $\vec{N}[n]$ can be stated as:

\begin{equation}
\vec{x}_[n] = \vec{\delta}[n]\vec{N}[n]
\end{equation}

\subsubsection{Statistical Features for Classification}
Both the aforementioned windowing strategies were independently employed for extracting $14$ statistical features. These features preserve the temporal information of the raw data by eliminating random noise wihtin the recorded signals. Such statistical features of windowing segments have been shown to provide a low-rank description of the dataset while preserving the temporal information. Therefore, they also don't incur high computational costs \cite{wang2004scalable}.
Table 3 lists and explains all the statistical features used for classification purposes.

\begin{table}[htbp]
	\caption{Set of $14$ statistical features extracted from each segment of the time-series}
	\vspace{0.1in}
	\label{Statistical Features}
	\centering
	\begin{adjustbox}{width=1\textwidth}
	
	\begin{tabular}{ccc}
			
			\toprule[1.3pt]
			\textbf{Feature} & \textbf{Mathematical Definition} & \textbf{Description }\\
			\midrule[1.3pt]
			
			Mean  & $\mu = \frac{\sum\limits_{i=1}^{n}x[i]}{n}$ & Average value of the data points within a window segment  \\ \\ 
			
			Standard Deviation  & $\sigma = \sqrt{\frac{\sum\limits_{i=1}^{n}(x[i]-\mu)}{n}}$ & Quantitative measure of spread of the data within a sampled window frame.  \\ \\
			
			Variance  & $\sigma^2 = \frac{\sum\limits_{i=1}^{n}(x[i]-\mu)}{n}$ & Measure of dispersion of data in a window, obtained by squaring the Standard deviation.  \\ \\
			
			Minimum Value  & $min (S) = x \in S \mid x \le y$ & Minimum value in a given window frame.
			\\ & $\forall y \in S$  \\ \\
			
			Maximum Value  & $max (S) = x \in S \mid x \ge y$ & Maximum value in a given window frame.
			\\ & $\forall y \in S$  \\ \\
			
			Jitter  & $\bar{J} = \frac{\sum\limits_{t=1}^{N}  \lVert x[t] - x[t-1] \rVert}{N-1}$ & Average amount of variation between adjacent samples of the data within a window frame.  \\ \\
			
			Mean Crossing Rate  & $X_n = \frac{\sum\limits_{t=1}^N \lVert I(x[t]>\bar{X})- I(x[t-1]>\bar{X}) \rVert}{N}$ &Rate at which adjacent data points cross the mean value of their respective windows. \\ & $where,$	$I(x)=
			\begin{cases}
			1 & $if $ x $ is true$\\
			0 & $otherwise$	
			\end{cases}$&  \\ \\
			
			Auto Correlation Mean  & $R_{xx}[K] = \frac{\sum\limits_{n=1}^{N-K} x[n]x[n+K] }{\sum\limits_{n=1}^{N-K}{x[N]}^2}$ &Measure of degree of relation between current sensor values and the future
			sensor values. \\
			& $and,$ $\bar{R}_{xx}= \frac{\sum\limits_{k=1}^N R_{xx}[K]}{N}$&  \\ \\
			
			Auto Correlation SD  & $\sigma_R = \sqrt{\frac{\sum\limits_{k=1}^N(R_{xx}[K]-\bar{R}_{xx})^2 }{N-1}}$ & Measure of the variation in the auto correlation values obtained by taking all possible \\
			&&lags of the raw data in the window currently being processed.\\ \\
			
			Auto Covariance Mean  & \scalebox{0.9}{$C_{xx}[k] = \sum\limits_{n=1}^{N-K}(x[n]-\bar{X})(x[n+K]-\bar{X})$}  & Measure of relationship between the current and the future values of the time series.\\ 
			& $and,$ $\bar{C}_{xx} = \frac{\sum\limits_{k=1}^NC_{xx}[K]}{N}$&\\\\
			
			Auto Covariance SD  & $\sigma_C = \sqrt{\frac{\sum \limits_{k=1}^N(C_{xx}[k]-\bar{C}_{xx})^2}{N-1}}$ & Measure of dispersion between the Auto Covariance values for different lags.\\ \\
			
			Skewness  & $SK_p = \frac{n}{(n-1)(n-2)} \sum\limits_{i=1}^N \frac{(X_i-\bar{X})^3}{\sigma^3}$ & Measure of amount of Symmetry in the data within a focused window.\\ \\
			
			Kurtosis  & $Kurt = \sum\limits_{i=1}^N \frac{(X_i-\bar{X})^4}{n\sigma^4}$ &  Feature used to provide description of tails of the distribution of data.\\\\ 
			
			Root Mean Squared Error  & $RMSE = \sqrt{\frac{1}{N} \sum\limits_{i=1}^N(X_i-\bar{X})^2}$ & Measure of error of individual data points relative to the average value within a window.\\ \\
			
			\bottomrule[1.3pt]
	\end{tabular}
\end{adjustbox}

\end{table}

\subsection{Classification Algorithms}
After feature extraction, we employed these features for training both BMI and age classifiers independently.
We used WEKA \cite{hall2009weka} for training and testing \textit{six} different classifiers. In our experiments, we use $10-fold$ testing for computing the true positive rates of all the classifiers. Below, we briefly highlight the classifiers used in BMI and age prediction experiments.

\subsubsection{J48 Decision Tree}
A J48 decision tree acts like a decision support system. It constructs a univariate decision tree from a dataset predicated upon the information gain by entropy of each attribute of the dataset \cite{bhargava2013decision}.  C4.5 algorithm \cite{quinlan2014c4} was used as a kernel for constructing a J48 tree. J48 decision tree has previously been proved to be useful for human activity classification from raw inertial sensor data in various studies \cite{walse2016study,wu2012classification,yuksek2011human,shoaib2013human} where Kundra \textit{et al}. \cite{kundra2014classification} have shown its utility in classifying electroencephalography (EEG) data.

\subsubsection{Multilayer Perceptron}
Multilayer Perceptron (MLP) belongs to the class of feedforward neural networks. MLP constructs a map between the inputs and output classes \cite{lakshmikantha2015human}. An MLP consists of three different types of layers i.e. An Input, Output and at least a single hidden layer where each layer in the network is fully connected to next. Each perceptron in any layer is activated according to sigmoid activation function given by:

\begin{equation}
y(v_i)=(1+e^{-v_i})^{-1}
\end{equation}

MLP trains the network using a supervised learning technique called Back Propagation. The weights on each perceptron are adjusted to minimize the error in the entire output, given by:

\begin{equation}
\varepsilon(n) = \frac{1}{2} \sum\limits_{j}e_j^2(n)
\end{equation}

The weights in the network are then adjusted to optimize the error, according to gradient descent algorithm as follows:
\begin{equation}
\Delta w_{ji}(n)= -\eta \frac{\delta\varepsilon(n)}{\delta v_j(n)} y_i(n)
\end{equation}

Where $\eta$ is the learning rate selected to be 0.3 in our analysis and the total number of layers in the MLP for each experiment were selected according to:

\begin{equation}
N_{layers} = \frac{1}{2}(N_{attributes}+N_{classes})
\end{equation}
Because of MLP’s popularity in classifying nonlinear data and its previous satisfactory results shown by \cite{li2016human,goh2014multilayer} in mining the raw inertial sensor data. MLP is particularly suited for our analyses. 
\\

\subsubsection{SVM}
The third classifier used was an SVM classifier using a polynomial kernel of degree 3. The SVM is a supervised learning classifier that classifies objects based on the support vectors of a dataset or points lie closest to the decision boundary. SVM maximize the distance between support vectors and the decision boundary \cite{wuhuman}. The objective function of SVM is given by:

\begin{equation}
min_{y,w,b} \qquad \frac{1}{2} \lVert w \rVert^2 + C \sum\limits_{i=1}^m \xi_i
\end{equation}

\begin{gather*}
s.t. \qquad y^(i)(w^Tx^(i)+b) \geq 1-\xi_i, i=1,2,...,m \\
\xi_i \geq 0, i=1,2,...,m
\end{gather*}

SVM has been found to be an outstanding classifier in cases having low training data instances and high dimensional features \cite{ruparel2013learning,forman2003extensive}. We demonstrate the accuracy of SVM on our dataset as the dimension of the feature vectors is increased.

\subsubsection{Random Forest}
A Random Forest classifier trains an ensemble of decision tree predictors on the training data and then uses a majority voting strategy on the results of each tree to classify the testing data. 

According to \cite{Breiman2001}, at each stage $k$ of the Random Forest Classifier, a random vector $\Theta_kis$ generated representing the random choices we make while generating the k’th tree. The tree at stage k is then a function of both the random data x chosen for its generation and the random vector $\Theta_k$ as:

\begin{equation}
h(x,\Theta_k)
\end{equation}

The Random Forest classifier is then represented mathematically by the notation:

\begin{equation}
\{h(x,\Theta_k) \qquad k=1,2,...\}
\end{equation}

\subsubsection{k-NN}
k-Nearest Neighbor or k-NN classifier belongs to the class of instance-based learning where each testing instance is compared with the training instances stored in memory rather than forming a general model of the whole data. This gives an additional advantage of classifying previously unseen data during training.

Given training data $(x_1,y_1 ),(x_2,y_2 ),...,(x_n,y_n),$ k-NN optimizes the Objective functions given as:

\begin{equation}
J = \sum\limits_{j=1}^{k}\sum\limits_{i=1}^{n} \lVert x_i^{(j)}-c_j\rVert^2
\end{equation}

\subsubsection{Logistic Regression}
Because of more than two classes, we employ multinomial logistic regression in our prediction towards Age and BMI classes. Multinomial regression extends the basic binary logistic regression by using a One vs All approach. The One vs All approach is as follows:

\begin{itemize}
	\item Train a Logistic regression classifier $h_\theta^(i)(x)$ for each class to predict the probability that $y = i$
	
	\item On a new input $x$, to make a prediction, pick the class $i$ that maximizes i.e.
\end{itemize}

\begin{equation*}
max_i h_\theta^{(i)}(x)
\end{equation*}

\section{Results and Discussion}
In this section, we present the analysis results from six different classification models, constructed for predicting human \textit{age} and \textit{BMI}. Prior to developing these schemes for conducting analyses, the dataset was subjected to preprocessing and feature selection (\textit{discussed earlier}). The process of extracting suitable, noise-free features, involved two different windowing functions--\textit{Gaussian }and \textit{Box }(or \textit{Square}) filters. Here, we present prediction results for both operations and hence compare their utility in classifying various subjects into age and BMI groups. In the end, we propose the most suitable model that could be leveraged for accurately predicting and classifying BMI and age from gait samples. Such schemes could then be used for monitoring physiological changes within the body.\par 

\begin{table*}[!htbp]
	\renewcommand{\arraystretch}{1.4}
	
	\caption{Changes in Classifier Accuracies ($\%$) owing to different Square Window sizes}
	\vspace{0.1in}
	\label{TABLE-3}
	\centering
	\begin{adjustbox}{width=1\textwidth}
	\begin{tabular}{cccccccc}
		\toprule[1.3pt]
		&\textsf{Width ($s$)} & \textsf{J48} & \textsf{Multilayer Perceptron} & \textsf{SVM} & \textsf{Random Forest} & \textsf{KNN} & \textsf{Logistic Regression} \\
		\midrule[1.3pt]
		
		& 0.17  & 73.41 & 76.72 & 83.48 & 85.68 & 86.84 & 66.00 \\
		& 0.33  & 81.74 & 89.28 & 89.10 & 92.24 & 93.90 & 71.58 \\
		& 0.50  & 81.29 & 91.03 & 91.42 & 92.58 & 93.48 & 72.19 \\
		BMI & 0.67  & 79.80 & 91.51 & 91.85 & 91.43 & 92.28 & 72.16 \\
		&\marktopleft{c1} \textbf{0.83}  & \textbf{81.18} & \textbf{94.62} & \textbf{91.94} & \textbf{93.76} & \textbf{92.37} & \textbf{74.84} \markbottomright{c1} \\
		& 1.00  & 76.30 & 91.07 & 88.96 & 90.69 & 89.95 & 72.46 \\
		&1.17  & 79.47 & 93.26 & 89.15 & 92.23 & 90.18 & 74.49 \\
		\hline 
		& 0.17  & 74.61 & 74.19 & 81.67 & 88.23 & 86.66 & 61.12 \\
		& 0.33  & 80.34 & 86.62 & 89.06 & 93.46 & 92.42 & 68.74 \\
		& 0.50  & 79.35 & 89.16 & 90.13 & 92.90 & 92.13 & 71.74 \\
		AGE & 0.67  & 80.65 & 90.66 & 89.90 & 93.80 & 91.68 & 72.58 \\
		& 0.83  & 80.97 & 93.76 & 92.04 & 94.84 & 90.86 & 78.60 \\
		& 1.00  & 79.53 & 89.95 & 89.95 & 91.69 & 87.59 & 75.19 \\
		&\marktopleft{c2} \textbf{1.11}	& \textbf{84.75}& \textbf{96.19}& \textbf{94.72}	& \textbf{94.87}& \textbf{90.18}& \textbf{81.38} \markbottomright{c2}\\	
		\bottomrule[1.3pt]
		
	\end{tabular}
\end{adjustbox}
	
\end{table*}

We selected the most optimal parameters (Width and SD respectively) for both Square and Gaussian filters by empirically computing True Positive Rates ($R_{TPR}$) of each classifier against different values of width and SD ($\sigma$).Table 4 lists the sizes (in \textit{seconds}) for different square windows, used for feature extraction. We varied the square window sizes from $0.17s$ to $1.17s$. The resulting features computed against these parameters were then used to compute classifier accuracies, tabulated against their respective window sizes (Table 4). Doing so, helped us identify optimal width parameter for designing the square window function. It can be observed that classifier accuracies increased with increasing the window size. In case of BMI prediction however, the classifiers performed optimally for a window size of $0.83s.$ For age classification, a similar behavior was observed at a window size of $1.11s$ (Table 4).\par 

On the same lines, different values of SD ($0.06s$ to $0.56s$) were employed towards determining optimal SD value for tailoring the Gaussian window function (Table 5). Classifier accuracies increased with an increase in SD (Table IV). For BMI classification the classifiers performed most accuractely at $\sigma = 0.36s$ At $\sigma = 0.5s$ and $\sigma = 0.56s$, the accuracies of all the classifiers tend to maximize for age classification, as can be seen in Table 5. Taken together, the aforementioned analysis helped us identify optimal parameters for designing the window operations. These parameters were then employed for constructing accurate models for predicting BMI and age groups.\par 

\begin{table*}[!htbp]
	\renewcommand{\arraystretch}{1.4}
	
	\caption{Changes in Classifier Accuracies ($\%$) owing to different Standard Deviations of Gaussian Filter}
	\vspace{0.1in}
	\label{TABLE-4}
	\centering
	\begin{adjustbox}{width=1\textwidth}
	\begin{tabular}{cccccccc}
		\toprule[1.3pt]
		&\textsf{Width ($s$)} & \textsf{J48} & \textsf{Multilayer Perceptron} & \textsf{SVM} & \textsf{Random Forest} & \textsf{KNN} & \textsf{Logistic Regression} \\
		\midrule[1.3pt]
		
		& 0.06  & 54.99 & 65.67 & 71.43 & 69.20 & 74.04 & 59.22 \\
		& 0.14  & 69.35 & 82.03 & 85.02 & 82.41 & 89.63 & 64.36 \\
		& 0.22  & 74.04 & 89.32 & 89.32 & 86.87 & 92.24 & 66.90 \\
		BMI & 0.31  & 84.64 & 93.16 & 92.17 & 93.93 & 95.78 & 77.11 \\
		&\marktopleft{c1} \textbf{0.36}  & \textbf{85.02} & \textbf{92.70} & \textbf{93.01} & \textbf{94.24} & \textbf{96.08} & \textbf{77.42} \markbottomright{c1} \\
		& 0.44  & 83.72 & 92.93 & 93.32 & 94.62 & 96.54 & 76.88 \\
		& 0.56  & 80.57 & 92.40 & 93.63 & 94.32 & 96.39 & 76.04 \\
		
		\hline 
		& 0.08  & 73.35 & 80.26 & 81.95 & 86.64 & 85.71 & 65.21 \\
		& 0.17  & 83.56 & 88.94 & 91.09 & 95.01 & 92.70 & 72.58 \\
		& 0.25  & 83.33 & 91.40 & 94.16 & 95.93 & 96.39 & 76.04 \\
		AGE & 0.33  & 83.64 & 89.78 & 94.39 & 96.24 & 97.31 & 78.42 \\
		& 0.42  & 82.33 & 89.86 & 94.24 & 96.47 & 97.31 & 77.50 \\
		& 0.50  & 83.56 & 91.17 & 94.32 & 96.62 & 97.24 & 79.72 \\
		&\marktopleft{c2} \textbf{0.56}  & \textbf{81.64} & \textbf{91.47} & \textbf{95.01} & \textbf{96.62} & \textbf{97.31} & \textbf{79.26}\markbottomright{c2}\\
		
		\bottomrule[1.3pt]
		
	\end{tabular}
	\end{adjustbox}
\end{table*}

Box and Gaussian filters are common pre-processing filters employed for time-series segmentation in non-stationary signal analysis. Box filter allows complete passage of information, inherently being an ideal filer. Gaussian windowing, on the other hand, is an important white Gaussian noise reduction filter. Therefore, to reveal the impact of these strategies on classifier accuracies, we performed a comparative analysis between Gaussian and Box filtering towards each classification task (Fig. 1 and 2, Fig. 3 and 4). \par 

\begin{figure}[h]
	\centering
	\includegraphics[width=0.9\linewidth]{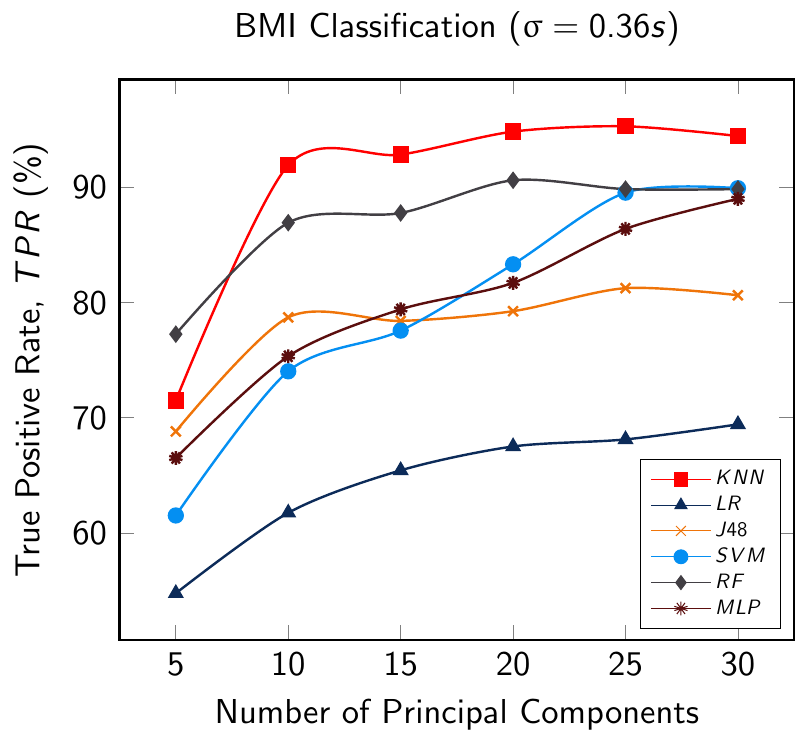}
	\caption{Gaussian BMI Prediction}
	\label{fig:figure-3-bmi-prediction-using-gaussian-windowing}
\end{figure}

Fig. 1 highlights the performance of different classifiers for classifying the subjects into various BMI groups. The dataset was processed using $\sigma = 0.36s$. From Fig.3, it can be seen that k-Nearest neighbour (KNN) performed the best with a $R_{TPR}$ of $94.47\%$. Moreover, random forest (RF), SVM and multilayer perceptron (MLP) performed with high accuracies of $89.89\%$, $89.94\%$ and $89.01\%$ respectively. Next, J48 decision trees performed with a $R_{TPR}$ of $80.64\%$ and logistic regression (LR) with a relatively low $R_{TPR}$ of $69.43\%$. Similarly, in case of square windowing operation we used $width = 0.83s$ (Fig.2). Here, KNN performed with the highest accuracy of $91.61\%$ followed by SVM ($90.86\%$), RF ($90.10\%$) and MLP ($89.24\%$) respectively.
All the classification tasks were performed at 10-fold testing to assure statistical significance of the results. Fig. 1 and 2 illustrate that $30$ principal components sufficed for attaining optimum classifier TPR.

\begin{figure}[h]
	\centering
	\includegraphics[width=0.9\linewidth]{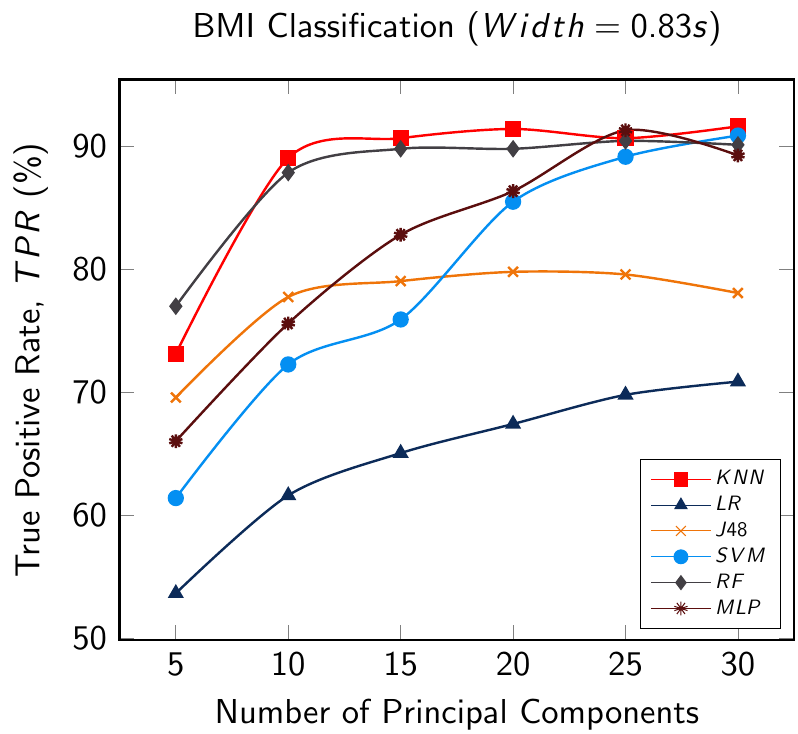}
	\caption{Square BMI Prediction}
	\label{fig:figure-4-bmi-prediction-using-square-windowing}
\end{figure} 

\begin{figure}[h]
	\centering
	\includegraphics[width=0.9\linewidth]{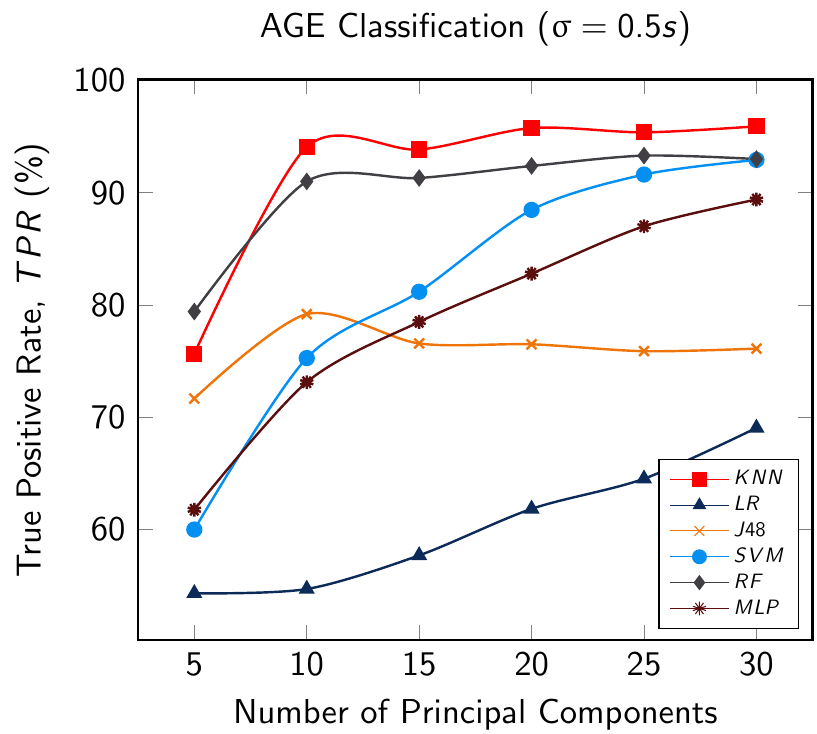}
	\caption{Gaussian Age Prediction}
	\label{fig:figure-5-age-prediction-using-gaussian-windowing}
\end{figure}

A similar protocol was repeated for classifying gait samples into various age groups (Fig. 3 and 4). The classifiers performed accurately for age classification as well. The dataset was processed at $\sigma = 0.5s$ and the classifier performance can be visualized in Fig.3. KNN performed most accurately ($95.9\%$) followed by random forest ($93.01\%$), SVM ($92.93\%$), MLP ($89.04\%$) and J48 ($76.11\%$). Logistic regression performed sub-optimally with a $R_{TPR}$ of $69.04\%$. For predicting age groups, we used square window size of $1.11s$ for extrating features. Performance assesment of Box filter has been highlighted in Fig. 6. The classification models performed suitably well for the age prediction task as well (Fig. 4). In this case, SVM classified subjects into various age groups with the highest accuracy of $94.28\%$. Multilayer perceptron (MLP ) and KNN, performed with an almost equivalent $R_{TPR}$ of $93.84\%$ and $93.69\%$ respectively. Random forest also performed well with an accuracy of $90.62\%$. J48 and logistic regression, however performed with low accuracies of $76.98\%$ and $73.167\%$ respectively.

\begin{figure}[h]
	\centering
	\includegraphics[width=0.9\linewidth]{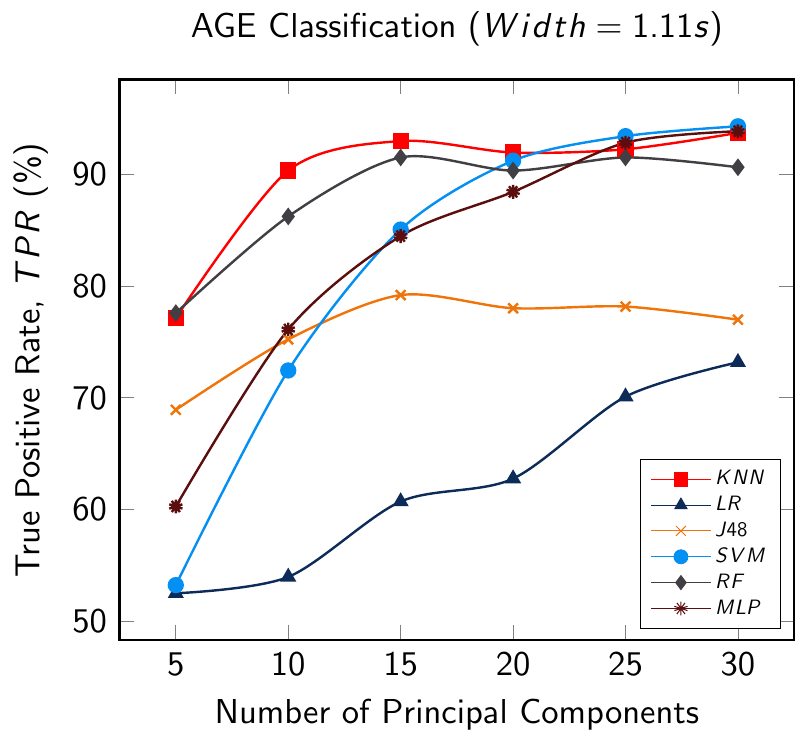}
	\caption{Square Age Prediction}
	\label{fig:figure-6-age-prediction-using-square-windowing}
\end{figure}

We now try to develop a mathematical intuition towards the enhanced performance of Gaussian filter over Square filter as the SD of Gaussian Filter is increased above a point, in our case, equal to $0.31s$ for BMI and $0.56s$ for Age classification. Figure () shows the Fourier transform of a typical square function, taking the shape of Sinc function. The Sinc function in frequency domain exhibits a continuum of lobes at various neighboring frequencies to the center frequency. This essentially means that sampling the non-stationary sensor data with a square filter has a undesired effect of sampling a spectrum of unwanted adjacent frequencies.

In case of a Gaussian Filter, as we see from Figure (), the frequency components are also of the Gaussian nature. However, on increasing the Standard deviation or informally, the spread of the Gaussian filter, the Gaussian Frequency response shrinks down, ultimately converging to a single peak in the frequency domain. At this Standard deviation, we presume that only one frequency component of the sensor data is being sampled by the Gaussian Filter. The features resulting from a single-frequency sampling of the raw data are hence less prone to noise and do not tend to over-fit the classifier with irrelevant information captured from adjacent frequencies as is the case with square filter. This explains the sudden rise in the accuracy of Gaussian Filter when the SD is increased above a point where its frequency response takes the form of a single point in frequency domain.

\section{Conclusions}
In this paper we have presented the first ever protocol to classify the Age and BMI of smart phone users using the raw Accelerometer and Gyroscope sensor data. We have also presented 14 statistical features extracted from the non-stationary smart phone sensor time-series and shared the performance of six common machine learning classifiers on these features. Furthermore, we described the use of two pre-processing filters to segment the raw time series into distinct frames and devised an empirical experiment to select the most optimal parameters of the two. In the end, we elaborate on the performance of each of the two sampling filters.





\bibliographystyle{model1-num-names}
\bibliography{./References}







\end{document}